# FPGA-Accelerated Real-Time Diagnostics at DIII-D Using the SLAC Neural Network Library for ML Inference

Abhilasha Dave[1], Semin Joung[2], SangKyeun Kim[3], Ramon Reed[3], Keith Erickson[3],
Jalal Butt[4], Azarakhsh Jalalvand[5], Mudit Mishra[1], James Russell[1], Larry Ruckman[1],
Ryan Herbst[1], Egemen Kolemen[3,5], David Smith[2], Ryan Coffee[1]

[1]SLAC National Accelerator Laboratory, Menlo Park, CA, USA
[2]University of Wisconsin, Madison, WI, USA
[3]Princeton Plasma Physics Laboratory, Princeton, NJ, USA
[4]Columbia University, New York, NY, USA [5]Princeton University, Princeton, NJ, USA

## 1 Abstract

In this work, we demonstrate the deployment of a hardware-accelerated machine learning (ML) inference system integrated into a real-time processing at the DIII-D tokamak fusion reactor. The team has successfully deployed an AMD/Xilinx KCU1500 field-programmable gate array (FPGA) into the realtime Plasma Control System (PCS) nodes that receives the live Beam Emission Spectroscopy (BES) signal used for Edge Localized Mode (ELM) forecasting. The FPGA hosts a dense neural network using the SLAC Neural Network Library (SNL) that has been trained to infer the likelihood of disruptive ELM conditions. This likelihood then feeds a separate plasma controller that uses Resonant Magnetic Perturbation coils to suppress the predicted disruptive condition. The SNL allows for on-the-fly updates of the neural network weights and biases without requiring full hardware resynthesis for the FPGA. Judicious design of the neural-network architecture can further allow for the hot-swapping of multiple classification tasks to be executed on the single FPGA, significantly enhancing the real-time adaptability of the system for context-aware control strategies that respond in real-time to evolving reactor conditions. These adaptive weights naturally support continuous model refinement and seamless task switching during live experimental operation.This use case is chosen as a high rate signal processing example that can serve as a template for general ML-based reactor diagnostic processing for active reactor control systems. We see this as an essential development for achieving reactor relevant operation in future continuous operation fusion devices.

## 2 Introduction

In this work, we present the design and deployment of a hardware-accelerated neural network model for real-time classification. We specifically target streaming diagnostic inference in support of plasma regime classification and breakthrough Edge Localized Mode prediction on the DIII-D tokamak reactor. The system enables dynamically reconfigurable inference running on Field Programmable Gate Array (FPGA) with 4.4 microsecond scale latency of the neural network inference. We detail the system architecture, the implementation of the ML model, the real-time integration with the Plasma Control System (PCS), and the performance benchmarks. These results highlight the potential for scalable, ML-based control frameworks in future reactor-relevant fusion environments and provide a template for extending to a broad range of autonomous control systems where inference latency and adaptability are both critical.

We chose chose magnetic confinement fusion as the demonstration case since it offers a promising pathway toward sustainable, carbon-neutral energy production, with the potential to meet rising global energy demands. Among the various reactor configurations under development, the tokamak is the technologically mature and shows strong potential for near-term commercialization [4]. Furthermore, the control system demonstrations that leverage the currently operating DIII-D facility are directly

applicable–if not in physical implementation then in functional replication–for other forms of magnetic confinement strategies. Achieving stable, high-performance operation in tokamaks remains a significant challenge due to the nonlinear and dynamic nature of confined plasmas. Real-time feedback control is therefore essential to regulate plasma stability, sustain confinement, and mitigate disruptions [18]. High-confinement regimes–H-mode–are critical for reaching reactor-relevant performance but are typically accompanied by edge-localized modes (ELMs) [7]. These ELMs pose significant risks to plasma stability and reactor component integrity. Real-time ELM prediction and mitigation are therefore vital to maintaining confinement and ensuring safe, reliable operation in future reactors. This is our motivation for using this real-time diagnostic signal processing case as our bench test for SLAC's Neural Library (SNL) for machine learning inference at the sensor edge.

Conventional plasma control systems, based on physics-informed models and empirical rules, often lack the adaptability needed to manage unanticipated transitions near operational limits [11]. This limitation has led to increasing interest in machine learning (ML) methods whereby the nonlinearity of the activation layers allows the model to learn complex nonlinear relationships from both high fidelity simulations and from high fidelity experimental data. These methods can provide classification, prediction, and even control in real-time plasma environments [3]. Machine learning models have shown promise in tasks such as disruption forecasting, tearing mode detection, and mode transition classification [18].

Deploying ML models within a real-time tokamak control system requires low-latency computational flows for live streaming diagnostic data. Several diagnostics at the DIII-D can support highsample rate signals: Electron Cyclotron Emission (ECE), $CO_2$ interferometry (CO2), and Beam Emission Spectroscopy (BES) the last of which offers particular sensitivity to disruptive ELM signals. This BES diagnostic provides high-resolution, two-dimensional poloidal (R,Z) measurements of electron density fluctuations with a 1 MSps digitization sample rate that captures signals of edge turbulence and transport dynamics. Real-time BES data are therefore used to classify plasma states and forecast breakthrough ELMs using a pre-trained neural network model. For real-time streaming signal processing and inference we have chosen to implement the model on an FPGA that is physically installed directly into the PCIe bus on the real-time signal acquisition node, RTSTAB.

The model is based on a multi-layer perceptron (MLP) architecture and is deployed using SLAC's SNL [10, 6, 16]. Inference is executed on the Xilinx FPGA [2] component in the RTSTAB node of the DIII-D real-time Plasma Control System (PCS)[11] which also includes an NVIDIA V100 GPU [13]. The neural network produces time-sequenced BES inference features used for regime classification and early ELM detection. This tight integration with the PCS ensures that control actions are executed within the stringent latency constraints required for effective real-time plasma control.

This work demonstrates the feasibility and scalability of deploying reconfigurable hardware-accelerated ML inference within real-time diagnostic pipelines. A key feature of the SNL framework is dynamic reloading of neural network parameters that allows task-specific weights and biases to be loaded at runtime without requiring full FPGA resynthesis. This enables the deployment of multiple inference tasks–such as ELM forecasting and confinement regime recognition–on a single firmware based neural network model. Our results establish a foundation for integrating adaptive and intelligent decisionmaking capabilities into critical control system settings such as next-generation plasma control systems, supporting disruption-resilient operation in future fusion reactors.

## 3  System Overview

The real-time plasma monitoring system deployed at the DIII-D tokamak integrates high-bandwidth diagnostics with the PCS in support of developing advanced control strategies. At the core of this architecture is a modular C++ based infrastructure that enables low-latency data acquisition, buffering, and inter-process communication in a heterogeneous compute environment [8, 9]. This infrastructure supports real-time data streaming from diagnostics with routing of the data streams to local acceleration hardware as in the FPGA used here. Further details of the specific system architecture and its implementation in the plasma control scenario will be presented in a forthcoming publication [11].

For the purposes of this manuscript, we briefly summarize the signal acquisition specific to the FPGA implementation in RTSTAB at DIII-D. Analog signals from the three front-end diagnostics, ECE, CO2, and BES, are digitized at 1 MSps rate for a total of 160 diagnostic channels with 18-bit differential signal Concurrent Real-Time (Concurrent-RT) [17] digitizers are housed in a PCIe-based RTSTAB acquisition server and synchronized via ring buffers. The architecture is designed to enable remote digitizers to appear as local devices to the host system, providing sub-microsecond latency across distances up to 100 meters [8, 9, 11].

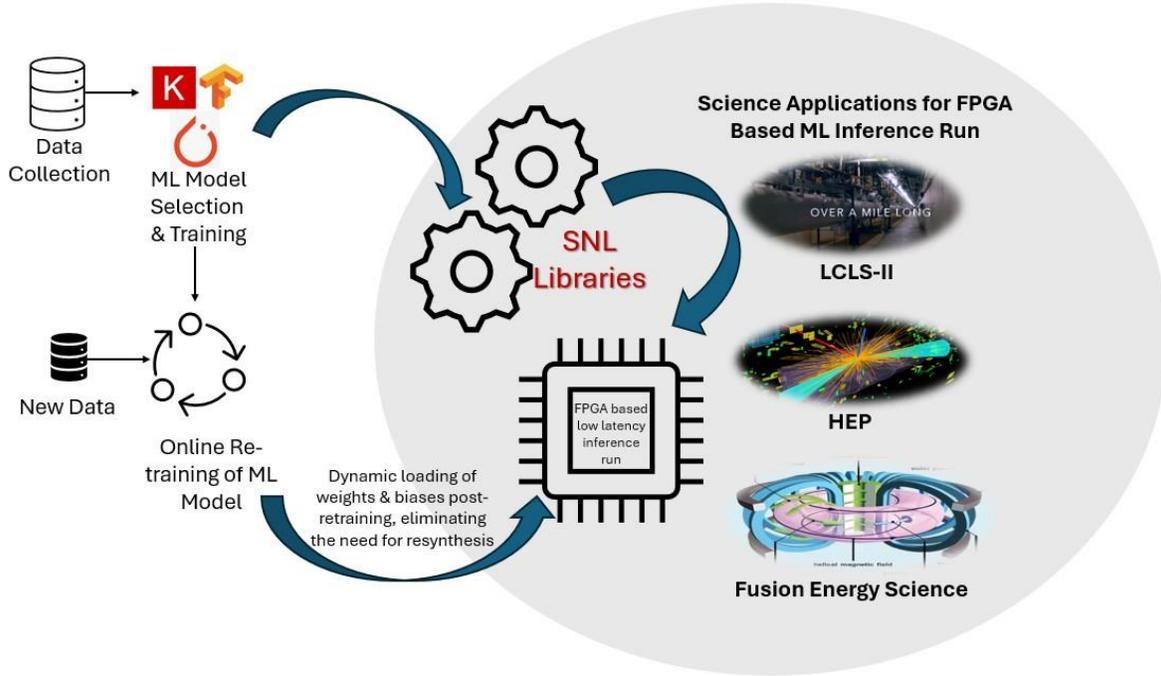

Figure 1: SNL High Level Design Flow reproduced from [6]

In this work, we have deployed the SNL-based machine learning inference model on the AMD/Xilinx KCU1500 FPGA. Processed outputs from the FPGA can be transmitted to the main PCS host via a high-speed InfiniBand interconnect. The InfiniBand network enabled the low-latency integration with the actuator control systems. This system architecture provides the deterministic timing, computational flexibility, and scalable bandwidth necessary for the deployment of intelligent, ML-enabled control systems in reactor-relevant fusion environments.

# 4 Application of the SLAC Neural Network Library for RealTime Plasma Monitoring in the DIII-D Tokamak

In this section, we introduce SNL, outline the motivation behind its development, and describe its integration within the DIII-D real-time plasma monitoring and control architecture.

It has become commonplace to use 6GS/s or higher digitization for time-series signals such as that used for the high repetition rate Linac Coherent Light Source II (LCLS-II) [1]. Such digitizers support a many tens of signal channels producing raw data streams reaching rates well over $100 GBs^{-1}$ [10]. In cases like the DIII-D reactor, however, the need for hundreds of digitized channels has thus far precluded all but the most recent upgrades of high rate diagnostics to sample rates of 1MS/s. The necessary processing of hundreds of signal channels in real time therefore presents significant challenges as it requires significant computation within limited latency and bandwidth budgets. Although machine learning (ML) methods offer promising avenues for accelerated data interpretation and reduction, inference pipelines running on CPUs or GPUs often introduce higher latencies than are compatible with the sub-millisecond timing constraints of real-time control systems such as at the DIII-D tokomak plasma reactor.

To address this challenge, SNL was developed as a high-performance, FPGA-based ML inference framework tailored for ultra-low-latency, high-throughput applications [10]. The SNL enables the deployment of ML models directly at the point of signal acquisition, potentially into the on-board FPGAs of the sensor devices themselves in order to reduce the burden on network and back-end systems. As shown in Figure 1, neural networks can be trained remotely on High Performance Computing (HPC) facilities using standard ML frameworks such as TensorFlow [12], Keras [5], or PyTorch [15]. The resulting trained models are subsequently compiled for FPGA deployment using SNL, a toolchain that has been developed using AMD Vitis, a unified software platform for AMD FPGA programming. This toolchain leverages high-level synthesis (HLS) and C++ templates to compile neural network models as FPGA-accelerated inference engines.

The SNL supports dynamic reloading of weights and biases without the time consuming re-synthesis or place and route cycles. This significantly reduce user interaction time and avoids common hardware compilation issues such as timing closure failures on FPGA. One can leverage this reloading of weights, in combination with input and output masking operations, to effectively reconfigure a network. By over-provisioning the inputs and output, as in this case we simply mask out the unused ECE and $CO_2$ input channels–though they are nevertheless fed into the FPGA–we can implement a new model that ingests more diagnostic signals or accommodates aging or failing sensors as is fairly commin in the radiation environments of nuclear reactors. Furthermore, given the advancements in continual learning [14], the ability to periodically update the weights and bias parameters in the model, holding the architecture fixed, will enable the system to adapt to evolving reactor conditions and remain robust to so-called concept drift as the mapping from signal to physics observable changes over time.

The SNL provides a Keras-like application programming interface (API) which lowers the adoption barrier for users familiar with typical software-based machine learning development. It supports deployment of neural networks with tens of thousands of parameters, subject to hardware resource availability, latency constraints, and the computational complexity of individual layers. It is optimized for a pipelined, streaming data-flow model and achieves end-to-end inference latencies ranging from just a few microseconds to milliseconds depending on the degree of parallelization. The exact latency depends on factors such as network depth, layer types (e.g., fully connected, convolutional, etc.), model size, as well as trade-offs made between FPGA resource utilization and real-time latency requirements. These capabilities make SNL particularly well suited for integration into low-latency feedback control systems in fusion reactors where fast continuous signal processing and computation is essential for plasma stability and performance.

Originally motivated by the microsecond scale data-flow needs of the LCLS-II [10], we integrated SNL into the DIII-D tokamak's real-time plasma monitoring and control infrastructure by deploying a user trained neural network model on a SLAC configured AMD KCU115 FPGA located within the RTSTAB compute node. The SNL-compiled neural network processes the incoming time-series BES signals, in real time, to classify confinement regimes and forecast breakthrough ELMs. The inferred ELM likelihood, $P$, is then routed to the broader PCS cluster over the infiniband network to a seperate controller model that adjusts the RMP coils for breakthrough ELM suppression. In this DIII-D implementation, SNL acts as an intermediary between the diagnostic front-end and the plasma control system, enabling rapid, FPGA-based decision making to support intelligent, disruption-resilient plasma operation. The modular, library-based architecture of SNL further supports application-specific extension and customization of internal components, enhancing its flexibility for diverse experimental control workflows.

### 4.1 Integration of Pre- and Post-Processing Units with SNL-Based Inference

The real-time hardware and software framework, SHIELD [8, 9], provides the interface between highbandwidth diagnostics and the PCS. To ensure compliance with the SHIELD framework within the DIII-D real-time plasma monitoring and control system, we integrated both pre- and post-processing units as interface boundaries that book-end the SNL-based neural network inference core. The complete FPGA data flow, including these components, is illustrated in Figure 2.

The pre-processor module is implemented on the same AMD KCU115 FPGA and is responsible for ingesting high-rate diagnostic data from the digitizer. The digitizer operates at a sampling rate of 1MS/s across 160 signal channels all of which are then moved to the FPGA. From this data stream, the FPGA-based pre-processor extracts the 16 specific BES channels and discards the non-BES signals for the purpose of this particular user pipeline.

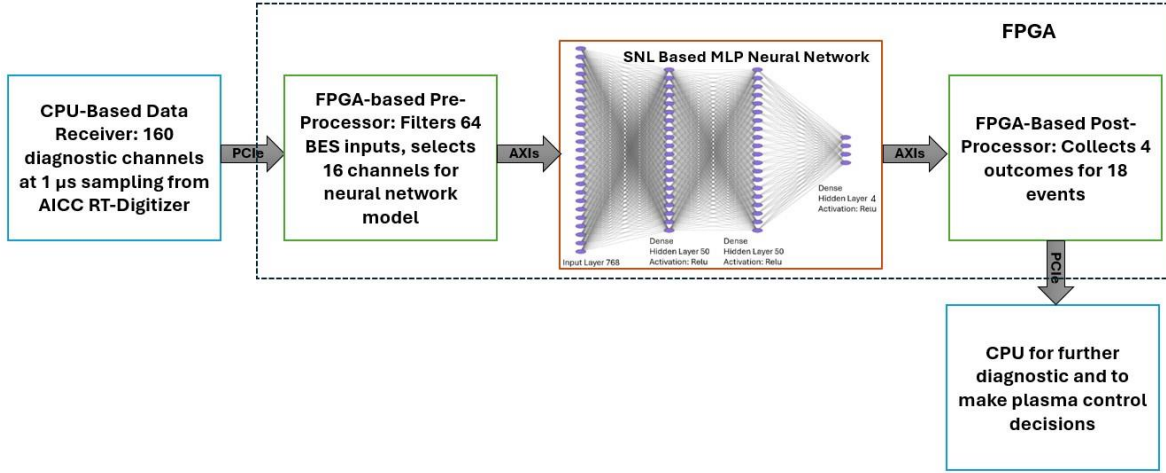

Figure 2: FPGA based Neural Network Design Data Flow

For each inference cycle, the system constructs a temporal input window comprising 48 consecutive time slices across the selected 16 BES channels. The neural network processes 18 such events per inference call, resulting in a total of 18 frames of 48 × 16 data points fed into the SNL-based neural network model. The SNL-compiled model performs a four-class classification and outputs one inference result per event. These classification outcomes are then passed to the post-processor, which aggregates and formats the results for downstream components. The final output is handed off via PCIe interconnect back to the host node and subsequently broadcast to the PCS where it is available for actuator control actions. This architecture enables low-latency, intelligent decision-making that aligns with SHIELD's modular, scalable vision for real-time experimental control, and demonstrates the feasibility dynamically reconfigurable ML inference in real-time fusion diagnostics.

## 4.2 Neural Network Structure

Both tasks, plasma confinement regime classification and breakthrough ELM event detection, use a shared neural network architecture built on a fully connected, feedforward topology. This model is compiled and deployed with SNL such that the architecture is intentionally lightweight to meet strict FPGA latency and resource constraints while retaining sufficient representational capacity for nonlinear classification tasks.

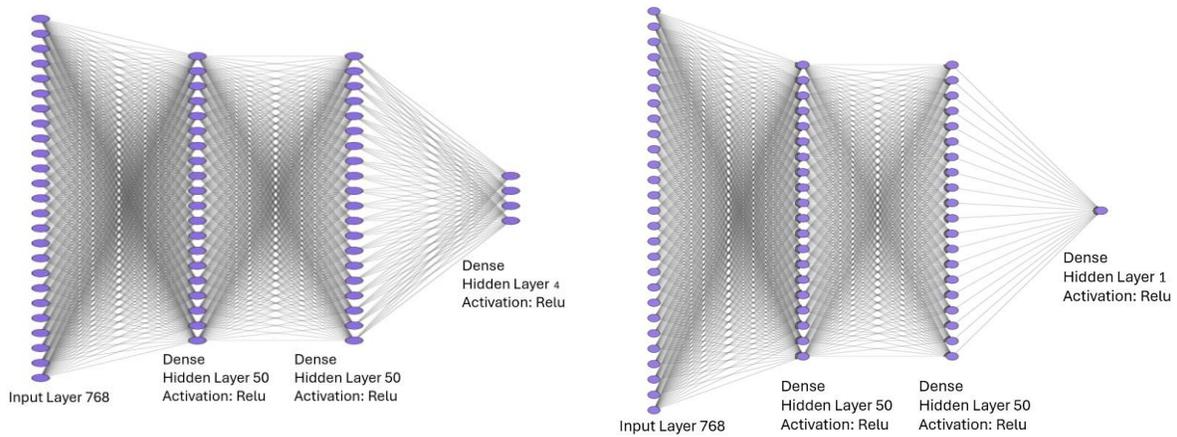

(a) Confinement Regime Classification: L-mode, H-mode, QH-mode, and wide pedestal QH-mode.

(b) ELM Event Classification.

Figure 3: Confinement and ELM classification figures shown side by side.

As illustrated in Figure 3a and Figure 3b, the network consists of the following components:
- **Input layer:** 768 features, corresponding to 48 consecutive time slices across 16 selected BES channels (i.e., a 48μs temporal observation window).

- **Two hidden layers:** Each comprising 50 neurons with ReLU activations to introduce nonlinearity.

- **Output layer:** Configurable with 1 to 4 neurons depending on the specific task. For example, a binary output for ELM detection or a 4-class output for mode classification (e.g., L-mode, standard H-mode, quiescent H-mode, and wide-pedestal QH-mode).

One trick in the design is our leveraging of the dynamic reconfiguration of the output layer. By leveraging SNL's support for runtime weight and bias reloading, the same FPGA-deployed model architecture can switch between a single output versus 4-class output without the need for resynthesis or recompilation. This enables even seamless task-switching of models for different plasma regimes or diagnostic scenarios, significantly improving flexibility while reducing system downtime.

To support high-throughput and low-latency operation, the FPGA implementation includes dedicated pre-processing and post-processing logic surrounding the neural network core. The pre-processor formats input data by transmitting eight features in parallel per clock cycle, significantly improving interface bandwidth. Rather than processing a single feature vector at a time, the pre-processor now aggregates and transmits batches of eight feature vectors simultaneously and it does it for it does it for 18 such input frames. For each inference cycle, it delivers a batch of 18 input frames, enabling the neural network to perform inference on all 18 events in a single operation. Similarly, the post processor aggregates the 4 outputs for the 18 frames and moves them back to the host as a block. On the output side, the post-processor collects the classification results, either scalar outputs or multi-class prediction vectors, for all 18 frames and packages them into a block structure compatible with the SHIELD software stack. Parallelizing across the 8 channels improves pipeline efficiency and aligns with the bandwidth and latency demands of the RTSTAB/PCS system. This batching strategy reduces the total inference time per 18 frame block and ensures that the neural network core operates at optimal utilization.

### 4.3 FPGA Implementation and Resource Utilization

The neural network model was deployed on the AMD KCU1500 development board which hosts a AMD Kintex-7 FPGA. Compiled using SNL from user-supplied trained neural network models that were originally developed and trained in PyTorch. The resulting compiled inference model is a fully pipelined hardware design optimized for ultra-low-latency inference. The final design, which includes the neural network inference engine was synthesized using Vitis HLS 2023.1. Table Table 1 summarizes the resource utilization for the complete design:

| **Neural Network** | **DSP %** | **FF %** | **LUT %** | **BRAM 18K %** | **Latency in** μs |
|---|---|---|---|---|---|
| 4 Class Classification Model | 5 | 8 | 13 | 1 | 5.28 |

Table 1: Vitis Synthesis report on FPFA resource utilization and latency report of SNL based MLP neural network model.

This utilization demonstrates sufficient headroom for deploying more complex models or supporting multiple inference tasks on a single FPGA. The design remains well within timing and routing constraints, achieving reliable operation at the target clock frequency of 6ns. During the recent closed loop ELM suppression experiment, the users executed a number of weight/bias reload cycles, thus exercising the model switching and adaptation for between-shot updates.

# 5    Future Work and System Integration Considerations

In the current implementation, the data ingestion pipeline is constrained by a CPU-based middleware layer that interfaces with the real-time digitizer. This software system acts as an intermediary between the digitizer hardware and the FPGA-based inference engine and delivers diagnostic data in fixed time windows. As a result, although the underlying FPGA architecture supports continuous streaming inference, the system is effectively driven in a segmented or "batched" format that is counter to the optimal data-flow paradigm for real-time inference. In future, one would more optimally design to avoid the software-mediated data transfer model and instead directly flow the signals first to the FPGA for inference and then output to the controller model, potentially also implemented in FPGA or other data-flow inference accelerator.

The current software-mediated design imposes a fixed temporal granularity on inference. That compromise of requires the neural network to process the 18 discrete input frames per inference cycle. While the FPGA is fully pipelined and optimized to handle such frame-based input efficiently, it limits the system's ability to exploit finer-grained or continuous and faster than real-time updates.

Future integration efforts will focus on direct digitizer to FPGA interfaces as is the case at the LCLSII [6, 1], thereby bypassing the intermediate CPU layer. Such improvements would enable continuous streaming inference, reduce control loop latency, and further optimize the ML inference pipeline to the microsecond-scale dynamics of plasma turbulence. Additionally, tighter coupling between data acquisition and hardware inference could support more adaptive or event-triggered control strategies–akin to high energy physics intelligent trigger systems–that could prove critical to advanced disruption avoidance and regime optimization in next-generation fusion reactors.

# 6    Conclusion

This work demonstrates the successful integration of a low-latency, FPGA-deployed neural network model into the DIII-D tokamak's real-time plasma monitoring and control infrastructure. Using the SLAC Neural Library, we deployed a fully connected feedforward model on an AMD KCU105 FPGA within the RTSTAB node of the Plasma Control System. The model enabled classification of confinement regimes and early detection of breakthrough ELM events by ingesting the high-resolution Beam Emission Spectroscopy diagnostic. The system achieves just over 5 microsecond inference latency, 10% of the ingestion time window, while interfacing seamlessly with the SHIELD framework and the Plasma Control System (PCS), supporting rapid actuator decisions essential for disruption avoidance.

To accommodate SHIELD-compliant operation, we implemented custom pre- and post-processing modules on the FPGA to interface with upstream digitized signals and to structure the model outputs for real-time use. Despite constraints imposed by CPU-based middle-ware in the current data acquisition pipeline, our architecture maintains strict latency budgets and demonstrates strong scalability for future deployment scenarios.

These results highlight the potential of hardware-accelerated, reconfigurable machine learning systems to support intelligent control in high-performance plasma regimes. This approach lays the foundation for more adaptive, data-driven control strategies in next-generation fusion devices where real-time diagnostic interpretation and rapid response will be essential for sustained and disruption-free operation.

# 7    Acknowledgments

This material is based upon work supported by the U.S. Department of Energy, Office of Science, Office of Fusion Energy Sciences, under Field Work Proposals 100636 "Machine Learning for Real-time Fusion Plasma Behavior Prediction and Manipulation" and 101046 "Unleash the Machine Learning control theoretical development on DIII-D." This work also leveraged work performed at the DIII-D National Fusion Facility, a DOE Office of Science user facility, under Award(s) DE-FC02-04ER54698, DE-SC0021275, and DE-SC0024527.

# 8 Disclaimer

*What is this? Ryan C is asking.* This report was prepared as an account of work sponsored by an agency of the United States Government. Neither the United States Government nor any agency thereof, nor any of their employees, makes any warranty, express or implied, or assumes any legal liability or responsibility for the accuracy, completeness, or usefulness of any information, apparatus, product, or process disclosed, or represents that its use would not infringe privately owned rights. Reference herein to any specific commercial product, process, or service by trade name, trademark, manufacturer, or otherwise does not necessarily constitute or imply its endorsement, recommendation, or favoring by the United States Government or any agency thereof. The views and opinions of authors expressed herein do not necessarily state or reflect those of the United States Government or any agency thereof.

# 9 Data Availability

All relevant DIII-D data supporting the findings of this study are available from the DIII-D National Fusion Facility, which is operated by General Atomics for the U.S. Department of Energy. Access to DIII-D data requires following the user protocols described on the DIII-D website. Specific questions regarding data availability can be directed to the corresponding author of this paper.

# References

[1] https://abaco.com/products/fmc134-fpga-mezzanine-card. url: https://abaco.com/products/fmc134-fpga-mezzanine-card.

[2] AMD. *FPGA KCU105 Card*. https://www.amd.com/en/products/adaptive-socs-andfpgas/evaluation-boards/kcu105.html. Accessed: 2025-02-08.

[3] Rushil Anirudh et al. "2022 review of data-driven plasma science". In: *IEEE Transactions on Plasma Science* 51.7 (2023), pp. 1750–1838.

[4] Richard J Buttery et al. "The advanced tokamak path to a compact net electric fusion pilot plant". In: *Nuclear Fusion* 61.4 (2021), p. 046028.

[5] Franc¸ois Chollet et al. *Keras*. https://keras.io. 2015.

[6] Abhilasha Dave et al. "FPGA-accelerated SpeckleNN with SNL for real-time X-ray single-particle imaging". In: *Frontiers in High Performance Computing* 3 (2025), p. 1520151.

[7] DIII-D National Fusion Facility. *2025 – Integrated control for access to and maintenance of Wide-Pedestal QH-mode*. https://d3dfusion.org/2025-12-08/. Accessed: April 22, 2026. Dec. 2025.

[8] K. Erickson. "Scalable Real-time Diagnostic Infrastructure Supporting Disruption Prediction and Avoidance". In: *24th IEEE Real Time Conference*. ICISE, Quy Nhon, Vietnam, Apr. 2024. url: https://indico.global/event/6805/contributions/58371/attachments/29468/52359/OS_Erickson_81.pdf.

[9] K. Erickson. "Scalable Real-time Framework Enabling Machine Learning Based Plasma Control". In: *IAEA Technical Meeting on CODAC, Data Management, and Remote Participation in Fusion Research*. Sao Paulo, Brazil, July 2024. url: https://conferences.iaea.org/event/377/ contributions/31677/.

[10] Ryan Herbst et al. "Implementation of a framework for deploying ai inference engines in fpgas". In: *Smoky Mountains Computational Sciences and Engineering Conference*. Springer. 2022, pp. 120–134.

[11] SangKyeun Kim et al. "Real-time plasma monitoring framework for advanced plasma control and ML-research in DIII-D". Unpublished manuscript. 2025.

[12] Mart´ın Abadi et al. *TensorFlow: Large-Scale Machine Learning on Heterogeneous Systems*. Software available from tensorflow.org. 2015. url: https://www.tensorflow.org/.

[13] NVIDIA. *V100 Card*. https://www.nvidia.com/en-gb/data-center/tesla-v100/. Accessed: 2025-02-08.

[14] Oleksiy Ostapenko et al. *Continual Learning with Foundation Models: An Empirical Study of Latent Replay*. 2022. arXiv: 2205.00329[cs.LG]. url: https://arxiv.org/abs/2205.00329.

[15] Adam Paszke et al. "PyTorch: An Imperative Style, High-Performance Deep Learning Library". In: *Advances in Neural Information Processing Systems 32*. Curran Associates, Inc., 2019, pp. 8024–8035. url: http://papers.neurips.cc/paper/9015-pytorch-an-imperative-style-highperformance-deep-learning-library.pdf.

[16] Hamza Ezzaoui Rahali et al. "Neural Network Acceleration on MPSoC board: Integrating SLAC's SNL, Rogue Software and Auto-SNL". In: *arXiv preprint arXiv:2508.21739* (2025).

[17] Concurrent Real-Time. *"The 64-Channel Analog Input Card."*. Accessed: 2025-11-24. 2025. url: https://concurrent-rt.com/products/hardware/real-time-i-o/analog/64-channelanalog-input-card/.

[18] Andrew Rothstein et al. "Enabling Integrated AI Control on DIII-D: A Control System Design with State-of-the-art Experiments". In: *arXiv preprint arXiv:2511.08818* (2025).